\documentclass[aps,twocolumn,amsmath,amssymb,tighten,floatfix,prb]{revtex4-1}

\usepackage{graphicx}% Include figure files
\usepackage{dcolumn} % Align table columns on decimal point
\usepackage{bm, bbm}      % bold math
\usepackage{curves}
\usepackage{epic}
\usepackage{amssymb}
\usepackage{wasysym}
\usepackage{epsf, times}
\usepackage{subfigure}
\usepackage{amsthm}

\newcommand{\beq}{\begin{equation}}
\newcommand{\eeq}{\end{equation}}
\newcommand{\bea}{\begin{eqnarray}}
\newcommand{\eea}{\end{eqnarray}}

\def\openone{\leavevmode\hbox{\small1\kern-4.2pt\normalsize1}}

\begin{document}

\title{Exact results on the quench dynamics of the entanglement entropy in the toric code}

\author{
Armin Rahmani
and 
Claudio Chamon
       }
\affiliation{
Physics Department, Boston University, Boston, MA 02215, USA
            }

\date{\today}

\begin{abstract}

We study quantum quenches in the two-dimensional Kitaev toric code model and compute exactly the time-dependent entanglement entropy of the non-equilibrium wave-function evolving from a paramagnetic initial state  with the toric code Hamiltonian. We find that the area law survives at all times. Adding disorder to the toric code couplings makes the entanglement entropy per unit boundary length saturate to disorder-independent values at long times and in the thermodynamic limit. There are order-one corrections to the area law from the corners in the subsystem boundary but the topological entropy remains zero at all times. We argue that breaking the integrability with a small magnetic field could change the area law to a volume scaling as expected of thermalized states but is not sufficient for forming topological entanglement due to the presence of an excess energy and a finite density of defects. 

\end{abstract}

\maketitle
%
%
%%%%%%%%%%%%%%%%%%%%%%%%%%%%%%%%%%%%%%%%%%%%%%%%%%%%%%%%%%%%%%%%%%%%%%%%%%%%%%

\section{introduction}\label{sec:introduction}

Cold atomic systems have provided an ideal playground for the experimental studies of the unitary evolution of thermally isolated quantum systems.~\cite{Greiner02,Sadler06,Bloch08} Apart from the experimental interest, studies of quantum evolution have shed light on some fundamental questions such as thermalization.~\cite{Rigol08,Linden09} 

A widely-studied scenario is the unitary evolution of a quantum state when the Hamiltonian is suddenly changed -- a ``quantum quench'',~\cite{Sengupta04,Zurek05, Polkovnikov05,Cherng06,Cazalilla06,Calabrese06} which has been the subject of many recent studies (see Refs.~[\onlinecite{Gristev07,Polkovnikov08,Sengupta08,Sotiriadis09}] for a few examples). A challenging question, which has been addressed theoretically in one dimensional systems, is the evolution of the entanglement entropy following the quench.~\cite{Calabrese07,Fagotti08} More generally, such studies are concerned with the fate of an out-of-equilibrium quantum state with a Hamiltonian that changes with some time-dependent protocol. 

The case where the coupling constants are changed across a quantum phase transition is of special interest. In recent experiments with spinor condensates, for example, the formation of ferromagnetic domains was observed in a quench to a Hamiltonian with a symmetry-breaking ferromagnetic ground state.~\cite{Sadler06} Even though the interactions after the quench favor a certain type of order, it is not at all obvious that the order should emerge out of equilibrium. Since symmetry-breaking order has a local order parameter, the non-equilibrium state can support the formation of ordered domains. But what about quenches across a topological phase transition? Would any of the topological characteristics of the equilibrium phase show up in the non-equilibrium wave-function?

Another interesting aspect of the non-equilibrium dynamics is the issue of thermalization and whether a large system can act as an effective heat bath for its subsystems. If thermalization occurs, due to the presence of some excess energy, the non-equilibrium state at long times should be in some sense \textit{close} to a thermal state at a finite temperature, which implies volume scaling for the entanglement entropy.

The qualitative discussions above motivate a detailed analysis of a quantum quench across a topological phase transition into the topological phase. We study the specific case of the Kitaev toric code model~\cite{Kitaev03} with spin-polarized initial states. A related quantum quench problem with topologically ordered initial states was recently studied numerically in Ref.~[\onlinecite{Tsomokos09}]. We focus on the time evolution of the entanglement entropy because it is intimately related to topological order through the topological entanglement entropy.~\cite{Kitaev06,Levin06} Entanglement entropy is also related to thermalization since, as mentioned above, volume scaling of the entropy as opposed to the area law is generically expected of thermal states.  

We compute exactly the bipartite entanglement entropy as a  function of time. This is an example where exact calculation can be done in 2+1D (thus far, analytical results for the time-evolution of entanglement had focused on 1+1D systems). We show that the entanglement entropy respects the area law at all times and if one adds disorder to the coupling constants, the recurrences or revivals are suppressed at long times and the entanglement entropy per unit ``area'' (boundary length) $P$ equals
\[S/P=\ln(4/e)\]
in the thermodynamic limit when the initial polarization is in the $x$ or $z$ direction.

We find the dependence of this saturation value on the initial state for an arbitrary direction of the initial magnetization. A new ingredient that comes in when considering these initial states is order-one corrections to the entanglement entropy from the convex and concave corners in the subsystem boundary. We find that the general form of the generated entanglement entropy is given by
\[S(t)=f (t)\: P+f_> (t)\: C_> + f_<(t)\: C_<\]
where $P$ is the perimeter of the subsystem and $C_>$ ($C_<$) the number of convex (concave) corners in its boundary. The functions of time $f$, $f_>$ and $f_<$ are independent of the subsystem geometry. From the form of the entanglement entropy above, we find that the topological entropy vanishes at all times.

We study the effects of breaking the integrability in a perturbative approach. We identify the mechanism by which an integrability-breaking magnetic field can change the area law to the volume law as expected of thermalized states. The physical picture is that the area law survives at low orders in an expansion in the integrability-breaking field. Going to higher orders however increases the thickness of a ring-like region around the subsystem boundary that contributes to the entanglement entropy. Eventually at high enough orders, the ring will collapse onto itself making volume scaling possible. In the perturbative treatment, we explicitly find that the topological entanglement entropy remains zero at low orders. We expect that the excess energy and the finite density of defects should prevent the formation of any topological order in a sudden quench even if one breaks the integrability.

The paper is organized as follows. In section~\ref{sec:quenchesXZ} we discuss quantum quenches for spin-polarized initial states in the $x$ or $z$ direction. We calculate the entanglement entropy analytically  and discuss the effects of disorder in the coupling constants. In section~\ref{sec:corners}, we study the role of the sharp corners in the subsystem boundary in the generation of entanglement by exactly calculating the second order Renyi entropy for an arbitrary direction of the initial magnetization. We then discuss the effects of breaking the integrability, which we treat perturbatively, in section~\ref{sec:non-integrable}. We conclude the paper in section~\ref{sec:conclusion}.

%%%%%%%%%%%%%%%%%%%%%%%%%%%%%%%%%%%%%%%%%%%%%%%%%%%%%%%%%%%%%%%%%%%%%%%%

\section{Integrable quench: initial magnetization in the $x$ or $z$ direction}
\label{sec:quenchesXZ}

Consider the toric code Hamiltonian in a field~\cite{Trebst07,Vidal09,Tupitsyn10}

\begin{equation}
H=- \sum_p \lambda_p B_p - \sum_s \mu_s A_s-h \sum_i\vec{n}\cdot\vec{{\sigma_i}}.
\end{equation}
The Hilbert space of the above Hamiltonian is given by spins living on the bonds of a square lattice on a torus and the sums are over all plaquettes, stars and bonds respectively. The plaquette and star operators are defined as follows
\[
B_p=\prod_{i \in p}\sigma^z_i, \qquad A_s=\prod_{i \in s}\sigma^x_i
\]
where $\sigma$s are the Pauli matrices. For the usual Kitaev model the coupling constants $\lambda$ and $\mu$ are independent of the plaquettes and the stars but here we keep the $p$ and $s$ indeces to allow for disorder in the couplings. Note that for the model in zero magnetic field ($h=0$), the ground state is independent of the actual values of the coupling constants $\lambda$ and $\mu$ as long as they are all positive.

One can study the dynamics of this model by changing the coefficients $\lambda$, $\mu$ and $h$ in time with some protocol. In this section, we consider a quantum quench where the system is initially in a fully magnetized state and then evolves with the integrable toric code Hamiltonian, namely, we are initially in the ground state for $h\neq0$ and $\lambda_p=\mu_s=0$ for all stars and plaquettes. We then turn off the field $h$ at $t=0$ (which does not change the quantum state) and turn on the toric code part of the Hamiltonian as follows:  $\lambda=\lambda(t)$ and $\mu=\mu(t)$ ($\lambda(0)=\mu(0)=0$).

Consider for now, the case where $\vec{n}=\hat{x}$ ($\vec{n}\cdot\vec{{\sigma}}=\sigma^x$). The ground state of the usual model (without disorder in the couplings) exhibits a topological phase transition: for $h>h_c$ the system is a paramagnet and for $h<h_c$, it is topologically ordered. This topological phase is robust against weak disorder in the couplings. Let us call the initial state $| \Omega\rangle$. In the case where the initial magnetization is in the $x$ direction $|\Omega_x\rangle=|11...\rangle$ in the $ |\sigma_1^x\sigma_2^x...\rangle$ basis. The initial state is a tensor product of single spin states and thus has no entanglement.

In the paramagnetic initial state we have $\langle \sigma^x\rangle=1$ for all spin operators. Before addressing the question of the entanglement entropy evolution, let us consider the simple problem of what happens to the $\sigma^x$ expectation value. As a function of time, we have $\langle \sigma^x(t) \rangle=\langle \Omega_x |U^{\dag} \sigma^x U| \Omega_x\rangle$ where $U$ is the evolution operator given by
\begin{equation}
 U=\exp \left(
i \sum_p \int_0^t \lambda_p (t^\prime) d t^\prime B_p
+i \sum_s \int_0^t \mu_s (t^\prime) d t^\prime A_s
\right).
\end{equation}
Since all star and plaquettes commute with one another, we have dropped the time ordering operator. If we consider a sudden quench, we have $\int_0^t\lambda_p(t^\prime)d t^\prime=\lambda_p t$ and $\int_0^t\mu_s(t^\prime)d t^\prime=\mu_s t$. Let us focus on sudden quenches for a more compact notation. At the end of the calculation we can replace $\lambda t$ and $\mu t$ with the corresponding integrals over time-dependent couplings for a more general protocol.

The expectation value of $\sigma^x_{ p p^\prime}$, a spin between two nearest-neighbor plaquettes $p$ and $p^\prime$ is then given by
\[
 \langle \sigma^x_{ p p^\prime}(t) \rangle=
 \langle \Omega_x |
 e^{-i t\lambda_{p} B_{p}} e^{-it\lambda_{p^\prime} B_{p^\prime}}\:
 \sigma^x_{ p p^\prime} \:
 e^{i t\lambda_{p} B_{p}}e^{i t\lambda_{p^\prime} B_{p^\prime}}
| \Omega_x\rangle.
\]
Since $B_p^2=\openone$, we have
\[e^{\pm i t\lambda_{p} B_{p}}=\cos(\lambda_{p} t)\pm i \sin(\lambda_{p} t)B_{p}.\]
We have $B_{p,p^\prime}\sigma^x_{ p p^\prime}=- \sigma^x_{ p p^\prime}B_{p,p^\prime}$. Therefore we get
\[
\langle \sigma^x_{ p p^\prime}(t) \rangle=
\cos\left(2 \lambda_{p} t \right) \cos\left(2 \lambda_{p^\prime} t \right).
\]
In the completely clean case (a theoretical abstraction), the average of $\langle \sigma^x_{ p p^\prime}(t) \rangle$ over time is equal to $\frac{1}{2}$ but any amount of disorder will eventually lead to a state which unlike the initial paramagnet has a vanishing time-averaged expectation value for $\sigma^x$. 

We next consider the main problem of this section, namely the evolution of the entanglement entropy. The density matrix of the system can be written as $\rho(t)=U|\Omega\rangle\langle \Omega|U^\dag$ where $U$ is the time evolution operator and $|\Omega\rangle$ is a generic initial state. By inserting the resolution of identity twice, we can write,
\begin{equation}
\rho(t)=\sum_{\alpha, \sigma}
\langle \alpha | U | \Omega \rangle
\langle \Omega | U^\dag | \sigma \rangle
| \alpha \rangle \langle \sigma |.
\end{equation}

We would like to calculate the entanglement entropy between two subsystems $A$ and $B$. Any star or plaquette operator acts on four spins which may all belong to $A$, all belong to $B$ or some belong to $A$ and some to $B$. In this last case, the star or plaquette operator must lie at the boundary of the two subsystems.
We can then write
\begin{equation}\label{eq:U}
U=U_A U_B U_\partial
\end{equation}
where $U_A$ and $U_B$ only act on spins in $A$ and $B$ respectively and $U_\partial$ only depends on star and plaquette operators at the boundary of the two subsystems. The three operators $U_A$, $U_B$ and $U_\partial$ commute with one another. Each basis state can be written as the tensor product of states for spins in $A$ and $B$,  $|\alpha\rangle=|\alpha_A\beta_B\rangle$. We now integrate out the spins in $B$ to find the reduced density matrix for subsystem $A$, $\rho_A(t)=\sum_{\beta_B}\langle \beta_B |\rho(t) |\beta_B\rangle$ and obtain,
\begin{equation}\label{eq:rho_A}
\rho_A(t)=\sum_{\alpha_A, \sigma_A, \beta_B}
\langle \alpha_A \beta_B | U | \Omega \rangle 
\langle \Omega | U^\dag | \sigma_A \beta_B \rangle
|\alpha_A \rangle \langle \sigma_A |.
\end{equation}
Notice that since the trace of an operator is independent of the basis, we can replace $|\beta_B\rangle$ by $U_B|\beta_B\rangle$ in the above expression. This amounts to calculating the trace in a rotated basis. We then substitute Eq.~(\ref{eq:U}) into Eq.~(\ref{eq:rho_A}) above and obtain
\begin{equation*}
\rho_A(t)=\sum_{\alpha_A, \sigma_A, \beta_B}
\langle \alpha_A \beta_B | U_A U_\partial|\Omega\rangle
\langle \Omega | U_A^\dag U_\partial^\dag | \sigma_A \beta_B \rangle
| \alpha_A \rangle \langle \sigma_A |.
\end{equation*}
The part of the Hamiltonian that only acts on $B$ spins has no effect on the time evolution of the reduced density matrix for subsystem $A$. In order to calculate the entanglement entropy between the two subsystems we need to find ${\rm tr}(\rho_A^n)$. We give a replica index to $\alpha$, $\beta$ and $\sigma$ and drop the subsystem index for brevity. States $|\alpha_i\rangle$ are in subsystem $A$ and $|\beta_i\rangle$ in $B$. Using $\langle \sigma_i|\alpha_{i+1}\rangle=\delta_{\sigma_i,\alpha_{i+1}}$ we obtain,
\begin{equation}
{\rm tr}(\rho_A^n)=
\sum_{\{\alpha_i\},\{\beta_i\}}
\prod_{i=1}^n
\langle \alpha_i \beta_i |U_A U_\partial | \Omega \rangle 
\langle \Omega | U_A^\dag U_\partial^\dag | \alpha_{i+1} \beta_i \rangle
\end{equation}
with $\alpha_{n+1}=\alpha_{1}$. Once again because traces are basis-independent we can replace $|\alpha_i\rangle$ by $U_A|\alpha_i\rangle$ in the above equation and we obtain an expression for ${\rm tr}(\rho_A^n)$ that only depends on $U_\partial$ and not the whole $U$. Let us define the following commuting
operators
\begin{equation}
U_p^{\pm}=e^ {\pm i \sum_{p \in \partial_p} \lambda_p B_p t},
\qquad
U_s^{\pm}=e^ {\pm i \sum_{s\in\partial_s} \mu_s A_s t}
\end{equation}
where $\partial_p$ ($\partial_s$) is the set of all boundary plaquettes (stars) which act on both $A$ and $B$ spins. We then obtain
\begin{equation}\label{eq:tr_rho_A_2}
{\rm tr} (\rho_A^n)=
\sum_{\{ \alpha_i \}, \{ \beta_i \}}
\prod_{i=1}^n 
\langle \alpha_i \beta_i | U^+_p U^+_s | \Omega \rangle
\langle \Omega | U^-_s U^-_p | \alpha_{i+1} \beta_i \rangle
\end{equation}
with $\alpha_{n+1}=\alpha_{1}$. We now focus on the simple case where initially all spins are aligned in the $x$ direction. In this case, $|\Omega\rangle=|\Omega_x\rangle$ is an eigenstate of $U_s$ and we can therefore substitute $U^+_s|\Omega_x\rangle\langle\Omega_x|U^-_s=|\Omega_x\rangle\langle\Omega_x|$ in Eq.~(\ref{eq:tr_rho_A_2}) above. Also $|\alpha\rangle$ and $|\beta\rangle$ are chosen in the $\sigma^x$ basis so the action of $B_p$ on a basis state is to flip the four spins around the plaquette $p$.  For concreteness, we assume that the partition $B$ contains all spins which lie inside a simply connected $M \times N$ rectangular region. We then have $2(M+N)-4$ boundary plaquettes. Let us consider the action of $U^{\pm}_p=\prod_{p\in\partial_p}e^{\lambda_p B_p t}$ on a state $|\alpha\beta\rangle$. Since $B_P^2=\openone$, we have
\begin{equation}\label{eq:Upm}
U^{\pm}_p=\prod_{p\in\partial_p}
\left(
\cos \lambda_p t \pm i B_p \sin\lambda_p t
\right).
\end{equation}

Notice that $\langle \alpha \beta |U^{+}_p|\Omega_x\rangle\neq 0$ only if the state the state $|\alpha \beta\rangle$ is obtained by acting with a certain number of $\partial_p$ plaquette flips on $ |\Omega_x\rangle$. Let us consider then Ising spin variables living at the center of  $\partial_p$ plaquettes. A plaquette is flipped for $\theta=-1$ and not flipped for $\theta=1$. Notice that, on a torus, flipping a certain set of plaquettes $\{\theta\}$  and flipping all other plaquettes that do not belong to $\{\theta\}$ have the same effect on the spins, i.e. they describe the same element in the group of plaquette flips. However, since we have  eliminated $U_A$ and $U_B$ from the formulation of the problem by a basis rotation, we can uniquely specify a state with nonvanishing $\langle \alpha \beta |U^{+}_p|\Omega_x\rangle$ by a set of flipped \textit{boundary} plaquettes. Let the state $|\alpha_1 \beta_1\rangle$ be labeled by $\{\theta_p \}$ for $p \in \partial_p$. If the partitions are not too small, it is not possible for a set of boundary plaquette flips acting on $|\Omega_x\rangle$ to create a state with the same $B$ spins and different $A$ spins or vice versa. Therefore $|\alpha_2 \beta_1\rangle=|\alpha_1 \beta_1\rangle=|\{\theta_p\}\rangle$ and we obtain using Eq.~(\ref{eq:tr_rho_A_2})
\begin{equation}\label{eq:trace-rho-n-x}
{\rm tr}(\rho_A^n)=
\sum_{\{\theta_p \},p \in \partial_p}
\left(
\:\langle \{\theta_p \} | U^+_p | \Omega_x \rangle 
\:\langle \Omega_x | U^-_p | \{\theta_p \} \rangle
\:\right)^n.
\end{equation}
Using Eq.~(\ref{eq:Upm}) we find that $\langle \{\theta_p \}|U^+_p|\Omega_x\rangle$ is a product of terms that are equal to $\cos\lambda_p t$ for each unflipped plaquette with $\theta_p=1$ and $i \sin\lambda_p t$ for each flipped plaquette with $\theta_p=-1$. We can then write
\begin{equation}\label{eq:matrix-element}
\langle \{\theta_p \} |U^+_p | \Omega_x \rangle =
\prod_{p\in\partial_p} 
\left(
\frac{1+\theta_p}{2} \cos \lambda_p t + i \frac{1-\theta_p}{2} \sin \lambda_p t
\right). 
\end{equation}
Switching the order of the sum and product after plugging Eq.~(\ref{eq:matrix-element}) into Eq.~(\ref{eq:trace-rho-n-x}), we obtain
\begin{equation*}
{\rm tr}(\rho_A^n)=
\prod_{p\in\partial_p}
\sum_{\theta_p}
\left(
\frac{1+\theta_p}{2} \cos^2 \lambda_p t + \frac{1-\theta_p}{2} \sin^2 \lambda_p t
\right)^n.
\end{equation*}
We have used the fact that for $\theta=\pm 1$, $\frac{1\pm\theta}{2}=\left(\frac{1\pm\theta}{2}\right)^2$.

Summing over $\theta_p=\pm 1$ for each boundary plaquette simply gives
\begin{equation}
{\rm tr}(\rho_A^n)=
\prod_{p \in \partial_p}
\left[
( \cos \lambda_p t )^{2n} + ( \sin \lambda_p t )^{2n} 
\right].
\end{equation}
By analytic continuation of $n$, we write the entanglement entropy as
\begin{equation*}
S = -{\rm tr} \left( \rho_A \log ( \rho_A ) \right)=
\lim_{n\rightarrow 1} \frac{\partial}{\partial n} {\rm tr}(\rho_A^n)
\end{equation*}
and we finally obtain,
\begin{equation}
    S=-\sum_{p \in \partial_p} \left[\cos^{2}\lambda_p t \ln
\left(\cos^{2}\lambda_p t\right)+ \sin^{2}\lambda_p t\ln \left(\sin^{2}\lambda_p
t\right)\right].
\end{equation}
For the clean Kitaev model, the entanglement entropy is an oscillatory function
of time whose maximum is proportional to the number of boundary plaquettes or
the perimeter of the subsystem. For this initial condition, the star term in the
toric code Hamiltonian has no effect on the evolution of the entanglement
entropy because the initial state is an eigenstate of that term. Similarly if initially all spins are in the $z$ direction, the plaquette term will have no effect. The completely clean system will never reach a steady state.

If there is some disorder in the coupling constants however, the entanglement entropy per unit boundary length will saturate to a steady state value in the thermodynamic limit. Suppose the number of boundary plaquettes $P$ is very large. As $t\rightarrow\infty$, due to the presence of some disorder in couplings $\lambda_p$, the phases $\lambda_p t$ are equally likely to take any value between $0$ and $2\pi$ modulo $2\pi$. Therefore, taking the thermodynamic limit first we can write $S(\{\phi_i\})=-\sum_{i=1}^P \zeta(\phi_i)$ with
\[
\zeta(\phi)\equiv \cos^{2}\phi \ln \left(\cos^{2}\phi\right)+
\sin^{2}\phi\ln \left(\sin^{2}\phi\right)
\]
where $\phi_i$ is a random variable between $0$ and $2 \pi$ drawn from a uniform distribution.
The average entanglement entropy is then given by
\begin{equation}
\bar{S}=-\frac{P}{2
\pi}\int_{0}^{2\pi} d\phi\:\zeta(\phi)=P \ln(4/e)
\end{equation}
We have $\overline{(S-\bar{S})^2}\propto P$ which implies that for any realization of disorder
\[
|\dfrac{S(\{\phi_i\})}{P}-\dfrac{\bar{S}}{P}|={\cal O}(\dfrac{1}{\sqrt{P}})
\]
indicating that in the thermodynamic limit, the entanglement per unit area length saturates to a steady-state value $\ln(4/e)$.
Notice that these results, namely the survival of area law at all times and the
oscillatory behavior in the entanglement entropy in the absence of disorder are generic features of integrable quenches where the integrals of motion are operators with local support. The entanglement entropy per unit boundary length relaxes in the thermodynamic limit to a steady state value in the presence of disorder in the coupling constants.

Because the system can reach a steady state, it is natural to ask if the steady state of a subsystem has the properties of a thermalized state described by a Gibbs or a generalized Gibbs ensemble.~\cite{Rigol07} For such thermalized states however, we expect the entanglement entropy of a small subsystem to scale as its volume in the limit of long times. The survival of the area law at all times which follows trivially from having local integrals of motion indicates the absence of thermalization even to a generalized Gibbs ensemble.

\section{Subsystem corners and the dynamical entanglement entropy}\label{sec:corners}

Let us now consider a more generic case where the initial density matrix is not
diagonal in either $\sigma^x$ or $\sigma^z$ basis and both terms in the Kitaev
Hamiltonian affect the entanglement dynamics. We find that even though in the 
thermodynamic limit the area law still applies, the physics of the out of
equilibrium entanglement is much richer. We find that here there are order-one corrections to the entanglement entropy from the concave and convex corners of the subsystem.

\begin{figure}
 \centering
 \includegraphics[width =5 cm]{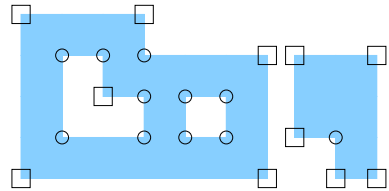}
 \caption{Subsystem B consists of spins inside the painted region, while subsystem A includes spins outside and on the boundary of this region. The convex (concave) corners for subsystem $B$ are represented by squares (circles).}
 \label{fig:corners}
\end{figure}

Subleading corrections to the entanglement entropy arising form the sharp corners in the boundary were found before for \textit{static} entanglement entropy in the logarithmic corrections to the area law in a class of conformal quantum critical points.~\cite{Cardy88,Fradkin06,Papanikolaou07} Here we find that corners can also make subleading contributions to the \textit{dynamical} entanglement. We find that when subsystem $B$ consists of spins inside a region with $C_>$ convex and $C_<$ concave corners as seen in Fig.~\ref{fig:corners}, the entanglement entropy after a quantum quench behaves as
\begin{equation}\label{eq:entropy-corners}
 S=fP+f_>C_> + f_< C_<
\end{equation}
where $P$ is the perimeter of the subsystem and $f$, $f_>$ and $f_<$ are functions of  time that do not depend on the subsystem geometry.
For a subsystem with a mostly smooth boundary and only a few sharp corners, the leading term is the $fP$ but for a subsystem with a corrugated boundary, the entanglement entropy can be different in the thermodynamic limit.

Notice that the number of concave and convex corners are not independent. Suppose subsystem $B$ consists of spins inside of an area with $m$ connected regions, each with a certain number of handles such that the total number handles is $g$. As an example in Fig.~\ref{fig:corners}, we have $C_>=C_<=11$ and $m=g=2$. We argue that 
\[
 C_>=C_<+4\:(m-g).
\]
This is because for a rectangular simply connected region, we have $C_>=4$ and $C_<=0$ and transforming the shape by adding or subtracting plaquettes on the square lattice does not change $ C_>-C_<$. Therefore each simply connected region makes a contribution of $+4$ to $C_>-C_<$ and by a similar argument, each hole contributes $-4$.
 \subsection{Initial magnetization in the $y$ direction}     
To see how the contribution from sharp corners comes in, consider an initial state with all the spins in the positive $y$ direction. Let us work in the the $\sigma^y$ basis with $\sigma^y|\pm\rangle=\pm|\pm\rangle$, $\sigma^x |\pm\rangle=|\mp\rangle$ and $\sigma^z|\pm\rangle=\mp i|\mp\rangle$. The action of a star or plaquette operator in this basis is then to flip the corresponding four $\sigma^y$ spins modulo a phase factor of $\pm 1$ from $B_p$. We will see later that this sign has no effect on the entanglement entropy. 

\begin{figure}
 \centering
 \includegraphics[width =7 cm]{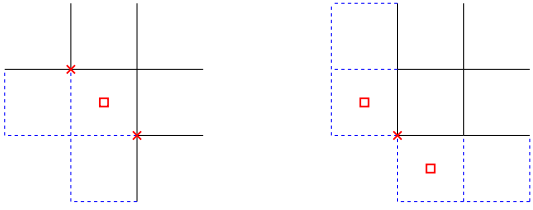}
 \caption{A convex (concave) corner of the $B$ subsystem is shown on the right (left) hand side. The dashed blue bonds belong to subsystem $B$ and the solid black bonds to subsystem $A$. A corner boundary plaquette (star) is represented by a red square (cross).}
 \label{fig:Ycorner}
\end{figure}

We assume for simplicity that there is a minimum distance of $3$ times the lattice spacing between all corners. Consider a convex corner as in the right hand side of Fig.~\ref{fig:Ycorner}. The $\beta$ spins live in the dashed blue (color online) bonds and the $\alpha$ spins on the solid black bonds. Flipping the corner plaquette represented by a red square has the same effect on $\beta$ spins as flipping the two boundary stars represented by red crosses. These operations of course have a different effect on $\alpha$ spins. Because we assumed a minimum distance between the corners, these corner star and plaquette operators uniquely correspond to one corner. Similarly as in the left hand side of Fig.~\ref{fig:Ycorner}, flipping the corner star at a concave corner has the same effect  on the $\alpha$ spins as flipping the two corresponding plaquettes. We see from this discussion that the effects of stars and plaquette flips mix at the corners, which leads to  a different entanglement generation than along a smooth boundary.

The details of the derivation will be given in appendix~\ref{App_A}. Let us write for now the result for the entanglement entropy for a  subsystem $B$ consisting of the bonds inside an $N\times M$ rectangular region. A boundary star or plaquette that is not located at a corner is refereed to as a regular star or plaquette. As seen in Fig.~\ref{fig:Ycorner}, at each convex corner, one plaquette and two stars are special. For our $N \times M$ subsystem, we have four convex corners, $2(M+N-4)$ regular boundary plaquettes and $2(M+N-6)$ regular boundary stars and the entanglement entropy for a clean system after a sudden quench is given by
\begin{equation}
 S=2(M+N-4)s_p+2(M+N-6)s_s+4s_c
\end{equation}
where the contributions from regular plaquettes ($s_p$), regular stars ($s_s$)
and convex corners ($s_c$) are given by
 \begin{eqnarray*}
s_p&=& (\sin^2 \lambda t \ln \sin^2 \lambda t+\cos^2 \lambda t \ln
\cos^2 \lambda t)\\
s_s&=& (\sin^2 \mu t \ln \sin^2 \mu t+\cos^2 \mu t \ln \cos^2 \mu t)\\
s_c&=& \left[r_1 \ln r_1+2r_2\ln r_2+r_3 \ln r_3\right] 
\end{eqnarray*} 
with
 \begin{eqnarray*}
r_1&=& \sin^2 \lambda t\sin^4 \mu t+\cos^2 \lambda t\cos^4 \mu t\\
r_2&=& \sin^2 \mu t\cos^2 \mu t\\
r_3&=& \sin^2 \lambda t\cos^4 \mu t+\cos^2 \lambda t\sin^4 \mu t.
\end{eqnarray*} 
The above solution conforms to the generic form Eq.~(\ref{eq:entropy-corners}) for
$P=2(M+N-2)$, $C_>=4$, $C_<=0$, $f=s_p+s_s$ and $f_>=s_c-8s_s-4s_p$.

In appendix~\ref{App_A}, we give a derivation of the above result in the more general setting of an arbitrary subsystem geometry. If instead of a sudden quench we turn on the coupling constants with some protocol $\lambda (t)$ and $\mu (t)$, we just need to replace $\lambda t$ and $\mu t$ by their integrals over time.

\begin{figure}
 \centering
 \includegraphics[width =7 cm]{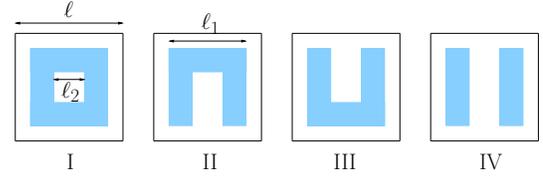}
 \caption{The partitions used for calculating the topological entanglement entropy.
}
 \label{fig:partitions}
\end{figure}

Using the partitions of Ref.~\onlinecite{Levin06}, shown in Fig.~\ref{fig:partitions}, the topological entanglement entropy is given by 
\begin{equation}
 S_{\rm topo}=S_{\rm I}-S_{\rm II}-S_{\rm III}+S_{\rm IV}.
\end{equation}
From the form of Eq.~(\ref{eq:entropy-corners}), we can then see that although there are order-one corrections to the area law, the topological entanglement entropy remains exactly zero at all times independently of how the toric code Hamiltonian is turned on.

It seems plausible that for \textit{any} initial state, as long as the integrability is not broken (with a field $h$ for example), evolution with the Kitaev Hamiltonian does not change the topological entanglement entropy for any time dependence of the coupling constants. This is because this integrable evolution does not allow propagation of information.
\subsection{Arbitrary initial magnetization}
The von Neumann entropy is a special case (limit of $n \rightarrow 1$) of the order $n$ Renyi entropy defined as 
\[
 S_n(\rho_A)=\frac{1}{1-n}{\rm tr}(\rho_A^n).
\]
Hereafter we refer to the second order ($n=2$) Renyi entropy as the Renyi entropy. The Renyi \ entropy $S_2(\rho_A)=-{\rm tr}(\rho_A^2)$ is simpler to calculate than the von Neumann entropy and is an equally valid measure of entanglement. It has been shown that the topological entanglement entropy calculated by any order Renyi entropy contains the same information in the abelean topological phases~\cite{Flammia09} such as the toric code. For an integrable quench from a paramagnetic state with magnetization in an arbitrary direction, we exactly calculate the Renyi entropy $S_2(\rho_A)$. We consider the clean case with position-independent coupling constants. With disorder in the couplings, the entropy saturates to the time-average of the clean case similarly to the simpler initial conditions considered before.

In the $\sigma_x$ basis, an arbitrary paramagnetic state can be written as
\[
 |\Omega\rangle=\left(c^+ |+\rangle+c^- |-\rangle \right) ^{\otimes N}.
\]
The calculation is rather involved and is done by a transfer matrix method explained in appendix~\ref{App_B}. We find the same general features for an arbitrary initial magnetization as for the initial magnetization in the $y$ direction. Let us summarize these features below.

 The Renyi entropy can be written as a sum of contributions from the perimeter of the subsystem as well as the convex and concave corners. These contributions are oscillatory functions of time in the absence  of disorder in the couplings which at long times, saturate to the values given by the time-average of the clean case if we have some disorder in the couplings. The saturation values are independent of the final coupling constants and the  protocol used for turning them on. They do however depend on the initial state.

\begin{figure}
 \centering
 \includegraphics[width =8 cm]{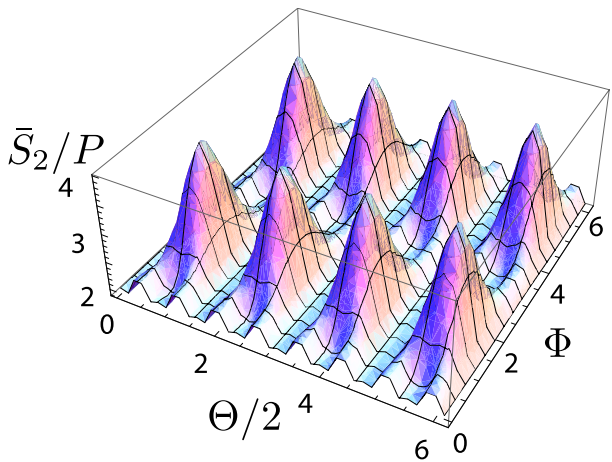}
 \caption{The dependence of the saturation value of the generated Renyi entropy per unit boundary length on the two angles parameterizing the magnetization direction of the initial state.} 
\label{fig:plot}
\end{figure}

For a large subsystem with only a few sharp corners, the leading term for the Renyi entropy is the first term in Eq.~(\ref{eq:entropy-corners}), namely $S\approx fP$. The saturation value of the Renyi entropy per unit boundary length is plotted in Fig.~\ref{fig:plot} as a function of the $\Theta/2$ and $\Phi$ angles which determine the initial state through $c^+=\cos \Theta/2\; e^{i \Phi}$ and $c^-=\sin \Theta/2$. Notice that this saturation value is a unique function of
$\langle B_p \rangle=\left(\sin  \Theta \cos \Phi \right) ^4$ and $\langle A_s \rangle=\left(\cos \Theta \right)^4$.  
 
\section{Breaking the integrability: entanglement entropy in the interaction picture}
\label{sec:non-integrable}

To study the effects of breaking the integrability, we focus on one simple case. Namely, we start with a state that has all spins aligned in the $x$ direction. As opposed to the integrable case where the magnetic field is turned off before turning on the integrable toric code Hamiltonian, the the toric code Hamiltonian is turned on in the presence of a small magnetic field $h$. We cannot perform exact analytical calculations in this case. We can however discuss the effects of the magnetic field by calculating a formal expansion for the (Renyi) entanglement and topological entropies in powers of $h$. The expansion can only be considered as a perturbative treatment if the toric code coupling constants are quickly pushed above $h$ and the time scales considered are not too large.

Two main insights are derived from this analysis. First we find indications as to how breaking the integrability can eventually lead to volume scaling in the entanglement entropy. The physical picture is that in the expansion in powers of $h$, the contribution to the entanglement entropy comes from ring-shaped areas around the subsystem boundary whose thickness grows at higher and higher orders in $h$. Eventually, at a high enough order, these correlated regions meet and the terms in the expansion no longer respect the area law. Even for a small $h$, these high order terms become important at long times making volume scaling of the entanglement entropy possible as expected of thermalized states. Secondly, we find that in this expansion in powers of $h$, the topological entanglement entropy remains zero at low orders until we reach an order proportional to the system size, suggesting that a small integrability-breaking term is not enough for generating topological entanglement.

A useful notion that we introduce in this section is the entanglement entropy in the interaction picture. We are used to the Shcr\"odinger, Heisenberg and interaction pictures for calculating the expectation values of observables, which are of course independent of the picture. If one defines the entanglement entropy as a property of the wave function alone, then since the wave function does not change in the Heisenberg picture, calculating the entanglement entropy in that picture leads to the nonsensical result that time evolution does not change the entanglement of a quantum state. An interesting alternative is to try and define the entanglement entropy entirely in terms of the correlation functions of appropriate operators. We do not pursue this path here. We argue instead that as explained below, the entanglement entropy calculated only from the wave function in the interaction picture can be a useful concept.

Let us first consider a \textit{non-interacting} Hamiltonian ${\cal H}_0$ such as a paramagnet. Consider a partitioning of the system into subsystems $A$ and $B$. The evolution operator for ${\cal H}_0$ can be written as
\[
{\cal U}_0(t)={{\cal U}_0}_A(t)\;{{\cal U}_0}_B(t)
\]
where ${\cal U}_A$ (${\cal U}_B$) only acts on the degrees of freedom in $A$ ($B$).
Notice that acting with ${\cal U}_0(t)$ on any wave function does not change its entanglement properties.
Now consider a Hamiltonian
\[
{\cal H}= {\cal H}_0+{\cal V}
\]
with the corresponding evolution operator ${\cal U}(t)$.

The wave functions at time $t$ in the Schr\"odinger and interaction pictures are respectively given by
\[|\psi_{\rm S}(t)\rangle={\cal U}(t)\:|\psi(0)\rangle,\qquad |\psi_{\rm I}(t)\rangle={\cal U}_0^\dagger(t)\:{\cal U}(t)\:|\psi(0)\rangle. \] 
Now since acting with ${\cal U}_0$ or  ${\cal U}_0^\dagger$ on any wave function does not change its entanglement, the interaction picture wave function has the same entanglement entropy as the Schr\"odinger picture. Note however the calculation
may be much simpler in the interaction picture.  

In the discussion above, we made use of the interaction picture entanglement entropy because
we had a Hamiltonian ${\cal H}_0$ whose evolution operator did not change the entanglement entropy. At the next level of complexity, where using the interaction picture wave function may be useful for calculating the entanglement entropy, the evolution operator of the \textit{unperturbed} Hamiltonian (${\cal U}_0$) may have another known property. For example it may only create entanglement entropy scaling with the subsystem area or it could be incapable of changing the topological entanglement. This is the case for the integrable toric code Hamiltonian. Now if we are interested in finding whether a perturbation can generate entanglement entropy with volume scaling or whether it can generate topological entanglement entropy, we can perform the calculation
for the wave function $|\psi_{\rm I}(t)\rangle$ in the interaction picture.

Let us go back to our quench problem and write the wave function in the interaction picture with
\[
 H_0(t)=- \sum_p \lambda(t) B_p - \sum_s \mu(t) A_s
\]
and 
\[V=h \sum_i{\sigma^x_i}.\]

Starting with the initial state $|0\rangle\equiv|\Omega_x\rangle$ with all the spins in the positive $x$ direction, the wave-function at time $t$ in the interaction picture is given by
\begin{equation}
 |\psi_{\rm I}(t)\rangle=e^{i\int_0^t dt^\prime H_0(t^\prime)}|\psi_{\rm S}(t)\rangle=\hat{U}_V(t)| 0\rangle
\end{equation}
where $\hat{U}_V(t)$ is given by
\begin{eqnarray*}
\hat{U}_V(t)&=&
1-ih\int_0^t d t_1 \hat{V}(t_1)\\
&+&(-ih)^2 \int_0^t dt_1 \int_0^{t_1} dt_2
\hat{V}(t_1)\hat{V}(t_2) +\cdots
\end{eqnarray*}
for
\[
\hat {V}(t)=h\;e^{-i\int_0^t dt^\prime \lambda(t^\prime)\sum_p B_p} \left(\sum_i \sigma^x_i \right) e^{i\int_0^t dt^\prime \lambda(t^\prime)\sum_p B_p}.
\]
For the simplicity of notation, we focus on the sudden quench and write $\int_0^t dt^\prime \lambda(t^\prime)$ as $\lambda t$, bearing in mind however that for a more general protocol, we can
substitute the integral at the end of the calculation.

We can explicitly calculate $\hat {V}(t)$ and the result is 
\begin{eqnarray}\label{eq:Vhat}
\hat{V}(t)&=&\frac{\cos 4 \lambda t +1}{2}\,C+\frac{\sin 4 \lambda t}{2}  \,P + \frac{\cos 4 \lambda t -1}{2} \,D
\nonumber\\ 
&\equiv& c(t)C+p(t)P+d(t)D 
\end{eqnarray}
with the operators $C$, $P$ and $D$ defined as follows
\begin{eqnarray}\label{eq:CPD_operators}
 C &=&\sum_{i} \sigma_i^x \nonumber \\
 P &=&\sum_{ p}\left(\;\begin{picture}(22,0)\put(2,-6){\includegraphics[width=0.035\textwidth]{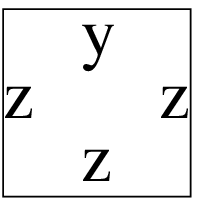}}\end{picture}+\begin{picture}(22,0)\put(2,-6){\includegraphics[width=0.035\textwidth]{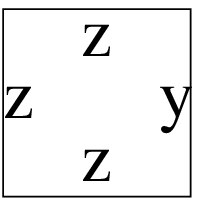}}\end{picture}+ \begin{picture}(22,0)\put(2,-6){\includegraphics[width=0.035\textwidth]{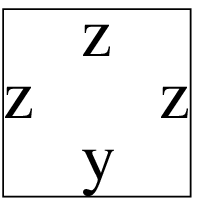}}\end{picture}+\begin{picture}(22,0)\put(2,-6){\includegraphics[width=0.035\textwidth]{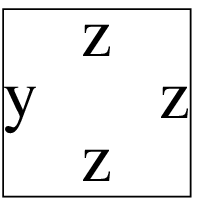}}\end{picture}\;\right)  \nonumber  \\
 D &=& \sum_{ d}\;\begin{picture}(22,0)\put(2,-6){\includegraphics[width=0.07\textwidth]{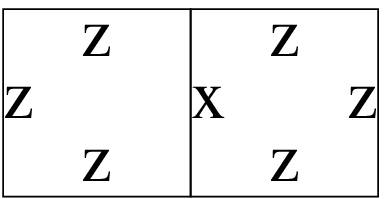}}\end{picture}\qquad.
\end{eqnarray}

The sums are over all bonds, plaquettes and dominos and the graphical notation represents a product of Pauli matrices. For example if the bonds in a plaquette are numbered clockwise from $1$ to $4$ with the top bond being number $1$ we have
\[
 \begin{picture}(22,0)\put(2,-6){\includegraphics[width=0.035\textwidth]{p1.eps}}\end{picture}=\sigma_y^1\sigma_z^2\sigma_z^3\sigma_z^4, \qquad \begin{picture}(22,0)\put(2,-6){\includegraphics[width=0.035\textwidth]{p2.eps}}\end{picture}=\sigma_z^1\sigma_y^2\sigma_z^3\sigma_z^4, \qquad \cdots
\]
The closed form in Eq.~(\ref{eq:Vhat}) is simply derived from the repeated substitution of
the following commutation relations
\begin{eqnarray*}
& &\left[\sum_p B_p,\;C \right]= 2 i P ,\qquad
\left[\sum_p B_p,\;P \right]= -4 i (C+D)\\
& &\left[\sum_p B_p,\;C+D \right]= 4 i P
\end{eqnarray*} 
in the Baker-Hausdorff expansion of $\hat{V}$
\begin{eqnarray*}
\hat {V}(t)&=&h\;\left(\sum_i \sigma^x_i+(-i \lambda t)\left[\sum_p B_p,\sum_i \sigma^x_i \right]\right.\\
  & &\left. +\frac{(-i \lambda t)^2}{2\:!}\left[\sum_p B_p,\left[\sum_p B_p,\sum_i \sigma^x_i \right]\right]+\cdots\right) 
\end{eqnarray*}
and resumming the series. 

The Renyi entropy for the interaction picture wave function $|\psi_{\rm I}(t)\rangle$ is given by
\begin{eqnarray}
 S_2(t)&=&-\log\: \langle \alpha_1 \beta_1|\hat{U}_V|0 \rangle\:\langle 0 |\hat{U}_V^\dagger|\alpha_2 \beta_1 \rangle\times\nonumber\\
 & &\langle \alpha_2 \beta_2|\hat{U}_V|0 \rangle\:\langle 0 |\hat{U}_V^\dagger|\alpha_1 \beta_2 \rangle
\end{eqnarray}
where summation over repeated indices is implied and as before, the $\alpha$ and $\beta$ spins
belong to the $A$ and $B$ subsystems respectively. 
First let us consider the first order term in the ${ \rm tr} (\rho_A^2)$ (the expression for the Renyi entropy without the log). We use the shorthand notation 
\[
\underbrace{VV\cdots V}_{ n\rm~times}=\int_0^t d t _1  \cdots \int_0^{t_{n-1}} d t _n \hat{V}(t_1)\cdots \hat{V}(t_n),
\]
and similarly for the Hermitian conjugate.
The first order term vanishes as seen below
\[
 { \rm tr} (\rho_A^2)=1-2 i h \langle 0|V|0 \rangle+2ih \langle 0|V^\dagger|0 \rangle +\cdots=1+O(h^2).
\]
The second order contribution to ${ \rm tr} (\rho_A^2)$ is given by
\begin{eqnarray*}
{ \rm tr} (\rho_A^2){\big |}_{o(h^2)}
&=&
h^2\;\left[\,2\,|\langle 0 \beta|V|0 \rangle|^2+2\,|\langle \alpha 0|V|0 \rangle|^2-\langle  0|V|0 \rangle^2\right.\\
 &-&\left. \langle  0|V^\dagger|0 \rangle^2-2\,\langle  0|VV|0 \rangle -2\,\langle  0|V^\dagger V^\dagger|0 \rangle\;\right].
\end{eqnarray*}

Let us assume that the whole system is on a torus and the total number of bonds
(spins) is $N_b$. The number of plaquettes $N_p=\frac{N_b}{2}$ and the number 
of dominos in the system is $N_d=N_b$. Let us introduce another shorthand notation
for integrals involving the functions $c$, $p$ and $d$,
\[
 (f_1,f_2,\cdots,f_n)=\int_0^t d t _1  \cdots \int_0^{t_{n-1}} d t _n f_1(t_1)\cdots f_n(t_n)
\]
so we have for example $(c)=\int_0^t d t _1 \:c(t_1)$ and $(c,p)=\int_0^t d t _1\int_0^{t_1} d t _2 \;c(t_1)\;p(t_2)$. We denote the number of plaquettes (dominos) with all bonds in subsystem $A$ by $n_{{p}A}$ ($n_{{d}A}$) and similarly by $n_{{p}B}$ ($n_{{d}B}$) for subsystem $B$.
We then have   
\begin{eqnarray*}
&|\langle 0 \beta|V|0 \rangle|^2={N_{b}}^2(c)^2+n_{{ p}B}\times 16\,(p)^2+n_{{d}B}(d)^2 \\
&|\langle \alpha 0 |V|0 \rangle|^2={N_{ b}}^2(c)^2+n_{{p}A}\times16\,(p)^2+n_{{d}A}(d)^2\\
&\langle  0|V|0 \rangle={N_{b}}(c)\\
&\langle  0|VV|0 \rangle={N_{b}}^2(c,c)+N_{p}\times 16 \,(p,p)+N_{d}(d,d).
\end{eqnarray*}

The factors of $16=-4i \times 4i$ come from the fact that each term in $P$ is a sum of four operators that flip the spins around a plaquette.
Noting that because $c(t_1)\:c(t_2)$ is obviously symmetric under the exchange $t_1\leftrightarrow t_2$,
we have $(c,c)=(c)^2/2$ and similarly for $(p,p)$ and $(d,d)$. We then obtain
\begin{eqnarray*}
 { \rm tr} (\rho_{A}^2){\big |}&=&1-h^2(N_p-n_{{p}A}-n_{{p}B})(p)^2\\
&-&h^2(N_d-n_{{d}A}-n_{{d}B})(d)^2,
\end{eqnarray*} 
which using $\log(1+x)=x+{\cal O}(x^2)$ gives
\begin{eqnarray}\label{eq:second-order}
 S_2&=&h^2(N_p-n_{{p}A}-n_{{p}B})(p)^2\\ \nonumber
&+& h^2(N_d-n_{{d}A}-n_{{d}B})(d)^2+{\cal O} (h^4).
\end{eqnarray}
This expression demonstrates the physical picture for building volume entanglement
which we alluded to before. We observe from the above expression that at this order
the contribution to the entropy is from a ring-shaped area of thickness two lattice
spacings around the subsystem boundary. At higher orders, the thickness of this correlated region, which contributes to the entanglement entropy, grows until the ring closes in on itself, making volume scaling possible.

Notice that in contrast to exact area law which is possible in a finite system, since the Renyi entropy is symmetric under exchanging the subsystems, volume scaling can only hold exactly for a finite subsystem in an \textit{infinite} system and as a good approximation in the limit when a subsystem is much smaller than the whole system.

It is a simple exercise in geometry to count $n_p$ and $n_d$ of the two subsystems for partitions shown in Fig.~\ref{fig:partitions} and verify directly that no topological entanglement is generated at order $h^2$. The details of this calculation will be given in appendix~\ref{App_C}. We will also show in that Appendix that at the next order with
a non-vanishing contribution to the Renyi entropy (${\cal O}(h^4)$), the topological term will still vanish due to an exact cancellation.

Similarly to the volume scaling, topological entanglement is non-perturbative and does not show up in an expansion in powers of $h$ in the thermodynamic limit. For a finite system, we expect non-vanishing contributions to the topological entropy to appear at orders that scale with the system size. Let us comment here that on physical grounds, in the case of a sudden quench, we do not expect any topological entanglement to form even at infinitely long times. A trend toward forming volume entanglement observed at low orders suggests that in the limit of large times, the generic thermalization behavior is expected. Although we cannot give a mathematical proof, the fragility of topological order to thermal fluctuations~\cite{Castelnovo07,Nussinov09} and the extensive excess energy in the system suggests that the non-equilibrium state, after a sudden quantum quench, will likely bear no signature of the equilibrium topological order of the phase.

We emphasize that the expansion used here cannot be treated as a perturbative treatment in the adiabatic limit where the toric code coupling constants are turned on slowly because in that case $\lambda<h$ for short times. We know that in the case of adiabatic evolution, we approximately remain in equilibrium, i.e. in the ground state manifold, and topological order emerges after a time of order system size.~\cite{Hamma08}  
\section{conclusions}\label{sec:conclusion}
We studied the quench dynamics of the entanglement entropy in the Kitaev toric code model. In the case of the integrable quench we found the entropy exactly as a function of time. The leading order contribution was found to scale as the subsystem perimeter with an oscillatory behavior that gets suppressed in the presence of some disorder in the coupling constants leading to the saturation of the entanglement entropy to disorder-independent values respecting the area law.

 We obtained results for the generated entanglement entropy as a function of an initial paramagnetic state with arbitrary magnetization. We studied the effect of the sharp corners in the subsystem boundary and found that in analogy with the static entanglement discussed in the literature, they make subleading contributions to the dynamically generated entanglement entropy as well. We showed that the topological entanglement entropy remains zero in this quantum quench. The results of the integrable quench as well as the analysis of the integrability-breaking perturbations strongly suggest that the non-equilibrium state after a quench into the topological phase does not develop any of the topological structure present in the equilibrium phase.

We used the concept of the entanglement entropy in the interaction picture which simplifies perturbative approaches to the calculation of the entanglement entropy. This allows us to tease apart in this particular problem the effects of the time evolution with an unperturbed Hamiltonian that cannot generate either entanglement with volume scaling or topological entanglement and focus only the part of the evolution that can possibly lead to volume scaling or topological entanglement. By studying the effects of an integrability-breaking perturbation, we found a mechanism where volume scaling of the generated entanglement entropy can emerge at high orders in perturbation theory as expected of thermalized states. The notion of thermalization after a quench naturally ties in with topological order through the fragility of topological order to thermal fluctuations.
\acknowledgements
We are grateful to C. Castelnovo, A. Hamma, M. Haque, A. Polkovnikov and M. Rigol for helpful discussions. This work was supported by in part by the DOE Grant DE-FG02-06ER46316.  

\appendix
\section{Derivation of entropy for $y$ direction initial magnetization}
\label{App_A}

We explicitly calculate in this appendix the contributions to the entanglement entropy from regular boundary stars and plaquettes as well as the convex and concave corners when the initial magnetization is in the $y$ direction.

First let us find the number of regular boundary stars ($S_r$) and plaquettes ($P_r$) in terms of the subsystem perimeter and the number of convex and concave corners. We define the perimeter $P$ as the total number of boundary plaquettes or boundary stars (they are equal). We then have
%%%%%%%%%%%%%%%%%%%%%%%%%%%%%%%%%%%%%%%%%%%%%%%%%
\[
 P_r=P-C_>-2C_<,\qquad S_r=P-2C_>-C_<.
\]
%%%%%%%%%%%%%%%%%%%%%%%%%%%%%%%%%%%%%%%%%%%%%%%%%

Once again we use Eq.~(\ref{eq:tr_rho_A_2}) which generally holds for any initial state $|\Omega\rangle$. The only intermediate states (in the $\sigma^y$ basis) which contribute to the ${\rm tr}(\rho_A^n)$ are the ones connected to $|\Omega_y\rangle$ by some star or plaquette flips. We label these states by variable $\theta=\pm1$ and $\psi=\pm 1$ living on the boundary plaquettes and stars respectively. A negative $\theta$ ($\psi$) indicates that the corresponding plaquettes (star) is flipped with respect to $|\Omega\rangle_y$.
%%%%%%%%%%%%%%%%%%%%%%%%%%%%%%%%%%%%%%%%%%%%%%%%%
\begin{equation}\label{eq:labeling}
|\alpha\beta\rangle=|\{\theta_{p_r} ,\psi_{s_r} ,\theta_{p_>},
\psi^j_{p_>} ,\psi_{s_<}, \theta^j_{s_<} \}\rangle
\end{equation}
%%%%%%%%%%%%%%%%%%%%%%%%%%%%%%%%%%%%%%%%%%%%%%%%%
where $p_r$, $s_r$, $p_>$ and $s_<$ label regular plaquettes, regular stars, convex corner plaquettes and concave corner stars respectively. For $j=1,2$, $\{\psi^j_{p_>} \}$ ($\{\theta^j_{s_<} \}$) are variables living on the two stars (plaquettes) corresponding to the convex (concave) corners. Notice unlike  regular star variables the two stars at a convex corner are labeled by the corner plaquette and similarly the two plaquettes at a concave corner are labeled by a corner star since a convex (concave) corner uniquely  corresponds to a corner plaquette (star).

Now suppose two such states have the same $\beta$ spins. All regular and concave corner plaquette and star flips must be the same for these two states. As seen in Fig.~\ref{fig:Ycorner} however, at a convex corner the $\theta$ or $\psi$ variables for these two states could be different if both the convex corner plaquette and the corresponding two corner stars simultaneously have opposite flips.  Then for two states $|\alpha \beta\rangle$ and $|\alpha^\prime \beta\rangle$ (which can be labeled as Eq. (~\ref{eq:labeling}) because they contribute ${\rm tr}(\rho_A^n)$) all their $\theta$ and $\psi$ variables must be the same except for the ones corresponding to convex corners, namely
$\{\theta_{p_>},\psi^j_{p_>}\}$ and $\{\theta_{p_>},\psi^j_{p_>}\}^\prime$
which follow the relation
\[
\{\theta_{p_>},\psi^j_{p_>}\}^\prime=
\{\gamma_{p_>}\theta_{p_>},\gamma_{p_>}\psi^j_{p_>} \}
\]
with auxiliary variable $\gamma_{p_>}=\pm 1$. Similarly we can introduce variables
$\rho_{s_<}=\pm 1$ for concave corner and for two states $|\alpha \beta\rangle$ and $|\alpha\beta^\prime\rangle$ we must have \[\{\psi_{s_<},\theta^j_{s_<}\}^\prime=\{\rho_{s_<}\psi_{s_<},\rho_{s_<}\theta^j_{s_<} \}.
\]

Introducing a replica index $l$ on $\gamma^l$ and $\rho^l$, we can write 
%%%%%%%%%%%%%%%%%%%%%%%%%%%%%%%%%%%%%%%%%%%%%%%%%
\begin{eqnarray*}
|\alpha_1\beta_1\rangle&=&|\{\theta_{p_r} ,\psi_{s_r} ,\theta_{p_>},
\psi^j_{p_>} ,\psi_{s_o}, \theta^j_{s_o} \}\rangle\\ 
|\alpha_2\beta_1\rangle&=&|\{\theta_{p_r} ,\psi_{s_r}
,\gamma^1_{p_>}\theta_{p_>},\gamma^1_{p_>}\psi^j_{p_>} ,\psi_{s_<},
\theta^j_{s_<} \}\rangle\\
|\alpha_2\beta_2\rangle&=&|\{\theta_{p_r} ,\psi_{s_r}
,\gamma^1_{p_>}\theta_{p_>},\gamma^1_{p_>}\psi^j_{p_>}
,\rho^1_{s_<}\psi_{s_<}, \rho^1_{s_<}\theta^j_{s_<} \}\rangle\\
|\alpha_3\beta_2\rangle&=&|\{\theta_{p_r} ,\psi_{s_r}
,\gamma^2_{p_>}\theta_{p_>},\gamma^2_{p_>}\psi^j_{p_>}
,\rho^1_{s_<}\psi_{s_<}, \rho^1_{s_<}\theta^j_{s_<} \}\rangle\\
\cdots
\end{eqnarray*}
%%%%%%%%%%%%%%%%%%%%%%%%%%%%%%%%%%%%%%%%%%%%%%%%%

Let us for the moment focus on the Renyi entropy $-{\rm tr} (\rho_A^2)$ for simplicity. The calculation for ${\rm tr} (\rho_A^n)$ will be very similar. First we can see from the form of the the states above that the sign factors from $B_p$ acting on states in the $\sigma^y$ basis have no effect. A $B_p$ plaquette flip can only give a different sign for the matrix elements involving $|\alpha_1 \beta_1\rangle$ and $|\alpha_2 \beta_1\rangle$ at a corner. But then the plaquette flip at that corner would also give a different sign for the matrix elements involving $|\alpha_1 \beta_2\rangle$ and $|\alpha_2 \beta_2\rangle$ and the overall sign for the product of the four matrix elements appearing in ${\rm tr} (\rho_A^2)$ will be positive.  

The calculation is then similar to the simpler cases with the initial magnetization
in the $x$ and $z$ directions except for each convex or concave corner there
is an additional auxiliary variable to sum over. Again, each matrix element can be written as a product similar to Eq.~(\ref{eq:matrix-element}). By simply bringing together terms we can write ${\rm tr} (\rho_A^2)$ in the following form:
%%%%%%%%%%%%%%%%%%%%%%%%%%%%%%%%%%%%%%%%%%%%%%%%%
\begin{equation}\label{eq:pre-factorization}
{\rm tr}(\rho_A^2)=\sum\prod_{p_r}f_{p_r}
\prod_{s_r}f_{s_r}\prod_{p_>}f_{p_>}\prod_{s_<}f
_{s_<}
\end{equation}
%%%%%%%%%%%%%%%%%%%%%%%%%%%%%%%%%%%%%%%%%%%%%%%%%
where the summation is over all $\theta$, $\psi$, $\gamma$ and $\rho$ variables for both replicas. We can take the sum Eq.~\ref{eq:pre-factorization} inside the products by switching their order and perform the sum for individual terms. This gives the factorized form below
\begin{equation}
{\rm tr}(\rho_A^2)=
\prod_{p_r}F_{p_r}\prod_{s_r}F_{s_r}\prod_{p_>}F_{p_>}\prod_{s_<}F_{s_<}.
\end{equation}
where we will compute the $F$s below. Assuming no disorder and taking the log of both sides, we immediately obtain our main result for the entropy
%%%%%%%%%%%%%%%%%%%%%%%%%%%%%%%%%%%%%%%%%%%%%%%%%
\begin{eqnarray*}
S&=&(P-C_>-2C_<)\ln F_{p_r}\\
&+&(P-2C_>-C_<)\ln F_{s_r}+C_>\ln F_{p_>}+C_<\ln F_{s_<}
\end{eqnarray*}
which conforms to the general expression Eq.~(\ref{eq:entropy-corners}).

For regular stars and plaquttes, the terms appearing in the above Eq.~(\ref{eq:pre-factorization}) are identical to what we obtained when the initial spins were in the $x$ or $z$ direction. For example
\[
f_{p_r}=\left(\frac{1+\theta_{p_r}}{2}\cos^2\lambda_{p_r} t+ \frac{1-\theta_{p_r}}{2}\sin^2\lambda_{p_r} t\right)^2.
\]

 Let us now define the following function for a more compact notation
\[
 g(\vartheta,x)\equiv\frac{1+\vartheta}{2}\cos^2 x+\frac{1-\vartheta}{2}\sin^2 x.
\]
We can then write
\begin{equation}
 f_{p_r}=\left[g(\theta_{p_r},\lambda_{p_r}t)\right]^2,\qquad f_{s_r}=\left[g(\psi_{s_r},\mu_{s_r}t)\right]^2.
\end{equation}

At the corners, we have a more complicated structure and for example, $f_{p_>}$, in addition to $\theta_{p_>}$, also depends on $\gamma^1_{p_>}$ and  $\psi^j_{p_>}$ for $j=1,2$.
Notice that for the Renyi entropy $\alpha_3=\alpha_1$ and therefore $\gamma^2_{p_>}=1$.
We can explicitly write
\begin{eqnarray*}
f_{p_>} &=&
g(\theta_{p_>},\lambda_{p_>}t)g(\gamma^1_{p_>}\theta_{p_>},\lambda_{p_>}t)\times\\
& &\prod_j g(\psi^j_{p_>},\mu^j_{p_>}t)g(\gamma^1_{p_>}\psi^j_{p_>},\mu^j_{p_>}t).
\end{eqnarray*}
It can be easily shown that
\[
 g(\vartheta,x)=|\sin x \cos x|\:e^{\vartheta \ln|\cot x|}.
\]
Using the above identity we can write
%%%%%%%%%%%%%%%%%%%%%%%%%%%%%%%%%%%%%%%%%%%%%%%%%
\begin{eqnarray*}
f_{p_>} & = & (\frac{1}{8}\sin 2\lambda_{p_>} t\: \sin 2\mu_{p_>}^1 t\:\sin
2\mu_{p_>}^2 t )^2 \times\\
& & e^{ (1+\gamma^1_{p_>})
\left(\theta_{p_>}\ln|\cot\lambda_{p_>} t|
+
\sum_j{\psi_{p_>}^j}\ln|\cot\mu_{p_>}^j t|
\right)}.
\end{eqnarray*}
A similar expression with $\theta\leftrightarrow\psi$,
$\lambda\leftrightarrow\mu$ and $\gamma\rightarrow\rho$ gives $f_{s_<}$.
We can now perform the summations and we obtain
\begin{equation}\label{eq:regular_F}
    F_{p_r}=\sum_{\theta_{p_r}}f_{p_r}=(\cos\lambda_{p_r} t)^{2n}+(\sin\lambda_{p_r} t)^{2n}
\end{equation}
with $n=2$ for the Renyi entropy and similarly for $F_{s_r}$ with $\lambda\leftrightarrow\mu$.
To calculate
\[
 F_{p_>}=\sum_{\theta_{p_>},\psi_{p_>}^1,\psi_{p_>}^2,\gamma^1_{p_>}}f_{p_>},
\]
we first perform the sum over $\theta$s and $\psi$s and then over
$\gamma$s and we obtain after some algebra
%%%%%%%%%%%%%%%%%%%%%%%%%%%%%%%%%%%%%%%%%%%%%%%%%
\begin{eqnarray}\label{eq:convex_F}
F_{p_>}&=&\sum_{k=1}^4{\big(}\frac{1}{\xi_k}\sin ^2 \lambda_{p_>} t\:
\sin ^2 \mu_{p_>}^1 t\:\sin ^2 \mu_{p_>}^2 t \\ 
    \nonumber &+&\xi_k\cos ^2 \lambda_{p_>} t\: \cos ^2 \mu_{p_>}^1 t\:\cos ^2
\mu_{p_>}^2 t {\big)}^n
\end{eqnarray}
%%%%%%%%%%%%%%%%%%%%%%%%%%%%%%%%%%%%%%%%%%%%%%%%%
with $n=2$ for the Renyi entropy and $\xi_1=1$, $\xi_2=\tan^2 \lambda_{p_>} t$, $\xi_3=\tan^2 \mu_{p_>}^1 t$ and
$\xi_4=\tan^2 \mu_{p_>}^2 t$. We can similarly find $F_{s_<}$ for
concave corners by switching $\lambda\leftrightarrow\mu$.
The generalization to $n>2$ is similar and leads to the appearance of a different $n$ in Eq.~(\ref{eq:convex_F}) and Eq.~(\ref{eq:regular_F}).

\section{Transfer matrix method for an arbitrary initial magnetization}
\label{App_B}
In this appendix, we discuss the transfer matrix method used in calculating the Renyi entropy for an arbitrary initial magnetization in the positive $\hat{n}$ direction. The initial state can be written as
\[
 |\Omega\rangle=\left(c^+ |+\rangle+c^- |-\rangle \right) ^{\otimes N}
\]
where $|+\rangle$ and $|-\rangle$ are the eigenstates of $\sigma^x$.

Let us begin by writing an explicit equation for $ {\rm tr}(\rho_A^2)$.
Using $n=2$ in Eq.~(\ref{eq:tr_rho_A_2}) we can write
%%%%%%%%%%%%%%%%%%%%%%%%%%%%%%%%%%%%%%%%%%%%%%%%%
\begin{eqnarray}\label{eq:Renyi-explicit}
{\rm tr}(\rho_A^2)&=&
\sum_{\alpha,\beta}
\langle \alpha_1 \beta_1 |U^+_pU^+_s| \Omega \rangle
\langle \Omega |U^-_sU^-_p| \alpha_2 \beta_1 \rangle\\ \nonumber
& & 
\langle \alpha_2 \beta_2 |U^+_pU^+_s| \Omega \rangle
\langle \Omega |U^-_sU^-_p| \alpha_1 \beta_2 \rangle.
\end{eqnarray}
%%%%%%%%%%%%%%%%%%%%%%%%%%%%%%%%%%%%%%%%%%%%%%%%%
Consider the first of the four matrix elements appeasing in the equation above, namely $\langle \alpha_1 \beta_1 |U^+_pU^+_s| \Omega \rangle$. We represent the $|\alpha_1 \beta_1\rangle$ spin configuration in a basis where boundary spins, i.e. spins belonging to a boundary star or plaquettes, are in the $\sigma_x$ basis and all other spins in the $\sigma_{\vec{n}}$ basis. In a schematic notation explained below, we can then write the matrix element as
\begin{eqnarray}
\label{eq:matrix-element-arbitrary}
\langle \alpha_1 \beta_1 |U^+_pU^+_s|\Omega\rangle
&=& \sum_{\{\theta\}} \left[
e^{i \mu t \sum \alpha^\prime \beta \alpha \alpha} \right .\\ \nonumber
& & 
\left. \left(\cos \lambda t\right)^{p_{\uparrow}}
\left( i\sin \lambda t\right)^{p_{\downarrow}}
{c^+}^{b_{\uparrow}}
{c^-}^{b_{\downarrow}} \right]
\end{eqnarray}
where in $\cos \lambda t$ and $\sin \lambda t$ we can more generally make the substitution $\lambda t \rightarrow \int_0^t \lambda(t^\prime)d t^\prime$.

\begin{figure}
 \centering
 \includegraphics[width =7 cm]{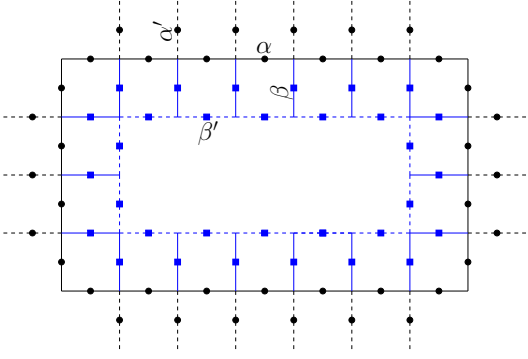}
 \caption{The notation used in Eq.~\ref{eq:matrix-element-arbitrary}. The boundary spins in subsystem $A$ are divided into $\alpha^\prime$ spins that do not belong
to any boundary plaquette and $\alpha$ spins that do belong to a boundary
plaquette. Similarly, the boundary spins in subsystem $B$
are decided into $\beta^\prime$ spins that do not belong
to any boundary star and $\beta$ spins that do belong to a boundary
star. We have represented the $\alpha^\prime$ spins by a black circle on a \textit{dashed}
black bond, the $\alpha$ spins by a black circle on a \textit{solid}
black bond, the $\beta^\prime$ spins by a blue square on a \textit{dashed}
blue bond and the $\beta$ spins by a blue square on a \textit{solid}
blue bond.}\label{fig:boundary_spins}
\end{figure}

The notation in the above equation needs some explanation. As before we have introduced Ising variables $\theta_p$ on the $P$ boundary plaquettes. For a given configuration of these variables $\{\theta\}$, $p_{\downarrow}$ and $p_{\uparrow}$ are the number of flipped and unflipped boundary plaquettes respectively and $b_{\downarrow}$ and $b_{\uparrow}$ are the number of up and down \textit{boundary} spins in the state obtained from $|\alpha_1\beta_1\rangle$ after flipping the plaquettes represented by $\{\theta\}$. We have introduced a new notation where $\alpha$ denotes only the boundary spins in subsystem $A$ that belong to a boundary plaquette and the other boundary spins in A, which only belong to a boundary star, are denoted by $\alpha^\prime$. Similarly boundary spins in subsystem $B$ that belong to a boundary star are labeled $\beta$ and the rest $\beta^\prime$ as seen in Fig.~\ref{fig:boundary_spins}. Assuming that the total number of boundary spins is $B$, we can relate $p_{\uparrow,\downarrow}$ and $b_{\uparrow,\downarrow}$ to the spins variables above as follows 
\[
p_{\uparrow}+p_{\downarrow}=P,
\qquad
p_{\uparrow}-p_{\downarrow}=\sum \theta
\]
and
\begin{eqnarray*}
b_{\uparrow}+b_{\downarrow} &=& B\\
b_{\uparrow}-b_{\downarrow} &=& 
\sum \alpha^\prime
+\sum \theta \alpha
+\sum \theta\beta^\prime
+\sum \theta \beta\theta.
\end{eqnarray*}  

In the sums above each $\alpha$ and $\beta^\prime$ spin is multiplied by the Ising variable $\theta$ of the boundary plaquette it belongs to. Each $\beta$ spin belongs to two boundary plaquettes and is therefore multiplied by the two corresponding $\theta$ variables. 
In the notation explained above, we now write the term $\langle \alpha_1 \beta_1 |U^+_pU^+_s|\Omega\rangle$
as follows
\begin{eqnarray*}
\langle \alpha_1 \beta_1 |U^+_pU^+_s|\Omega\rangle
&=&
(c^+ c^-)^{\frac{B}{2}}\:
(i\sin\lambda t\cos\lambda t)^{\frac{P}{2}}\:
e^{i \mu t \sum \alpha^\prime \beta \alpha\alpha}\:
\\
&\times & A^{\frac{1}{2}\sum \alpha^\prime}\sum_{\{\theta\}}A^{\frac{1}{2}(\sum \theta \alpha+\sum \theta
\beta^\prime+\sum \theta\beta\theta)}
\:H^{\frac{1}{2}\sum \theta}
\end{eqnarray*}
where 
\[H\equiv-i\cot \lambda t, \qquad A\equiv \frac{c^+}{c^-}.\]

Like the one-dimensional classical Ising model in a magnetic field, we can now integrate out the $\theta$ Ising variables using the transfer matrix method. The terms containing $\theta$
are promoted to $2\times2$ transfer matrices where each transfer matrix $T_{2\times2}$
is between two adjacent boundary plaquettes and
\begin{equation}\label{eq:two-by-two}
\langle \alpha_1 \beta_1 |U^+_pU^+_s|\Omega\rangle
=(c^+ c^-)^{\frac{B}{2}}\:
(i\sin\lambda t\cos\lambda t)^{\frac{P}{2}}\:
{\rm tr}\left( \prod T_{2\times2}\right).
\end{equation}
The other degrees of freedom can be systematically integrated out by repeatedly using the following trace identity
\begin{equation}\label{eq:identity}
{\rm tr}(A_1 A_2 \cdots)\:
{\rm tr}(B_1 B_2 \cdots)
={\rm tr}\left[\:(A_1\otimes B_1)\:(A_2\otimes B_2)\cdots\right].  
\end{equation}
Repeated use of the identity Eq.~(\ref{eq:identity}) leads to enlarged transfer matrices as discussed below. Tracing out all degrees of freedom will lead to $64\times64$ transfer matrices the trace of whose product gives the Renyi entropy.

Notice that at concave and convex corners, we get a different $2\times2$ transfer matrix. The $2\times2$ transfer matrices at the corners will be given at the end of this appendix for completeness. We discuss below the procedure for obtaining the final $64\times64$ transfer matrices from the $2\times2$ ones along a smooth boundary. At the sharp corners, we can obtain the $64\times64$ transfer matrices from the corresponding convex and concave corner $2\times2$ matrices essentially in the same way.

\begin{figure}
 \centering
 \includegraphics[width =8 cm]{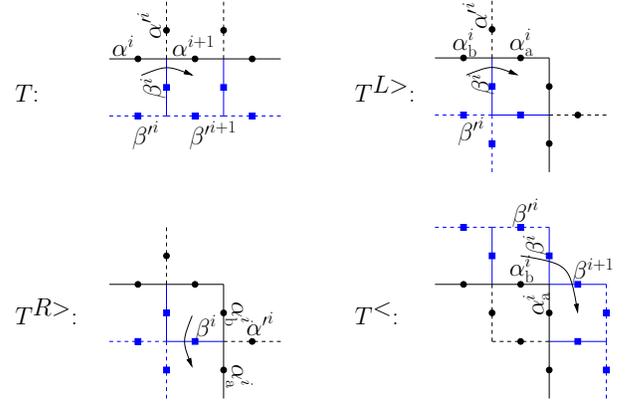}
 \caption{Top left: the transfer matrix $T_{2\times2}$ along a smooth boundary and the corresponding spin variables. Top right: the transfer matrix $T^{L>}_{2\times2}$ for entering a convex corner. Bottom left: the transfer matrix $T^{R>}_{2\times2}$ for exiting a convex corner. Bottom right: the transfer matrix $T^{<}_{2\times2}$ for a concave corner.}
 \label{fig:transfer_all}
\end{figure}

Let us now explicitly write $T_{2\times2}$ in Eq.~(\ref{eq:two-by-two}) along a smooth boundary. For a bond $\beta^i$ between two adjacent boundary plaquettes as show in the top left corner of Fig.~\ref{fig:transfer_all}, we have 

\begin{eqnarray*}
\lefteqn{
T_{2\times2}=
A^ {\frac{1}{2} {\alpha^\prime}^i }
\exp \left[{i \mu t {\alpha^\prime}^i \beta^i \alpha^i \alpha^{i+1} }\right]
\times} \\
& &
\left( \begin{array}{ll}
 H^{\frac{1}{2}}A^{\frac{1}{2}({\beta^\prime}^i+\alpha^i)}&0  \\ 
 0&H^{-\frac{1}{2}}A^{-\frac{1}{2}({\beta^\prime}^i+\alpha^i)}
\end{array}
\right)
\left(\begin{array}{ll}
 A^{\frac{1}{2}\beta^i}&A^{-\frac{1}{2}\beta^i}  \\ 
 A^{-\frac{1}{2}\beta^i}&A^{\frac{1}{2}\beta^i}
\end{array}
\right).
\end{eqnarray*}

The next step is to consider
\begin{equation*}
R_{1,2}\equiv \sum_{\beta_1}
\langle \alpha_1 \beta_1 | U^+_pU^+_s | \Omega\rangle
\langle \Omega | U^-_sU^-_p | \alpha_2 \beta_1 \rangle.
\end{equation*}
We can now use the trace identity Eq.~\ref{eq:identity} to bring together terms with the same $\beta$ and $\beta^\prime$ and trace out these degrees of freedom to get 
\begin{equation*} 
R_{1,2}=
(|c^+ c^-|)^{B} (\sin \lambda t \cos\lambda t)^{P} 
{\rm tr} \left( \prod T_{4\times4} \right)
\end{equation*}
where the $4\times 4$ transfer matrix above is given by
\begin{equation*}
T_{4\times4} = \exp
\left[ i\: \beta\:\mu t
\left (
{ \alpha^\prime }^i_1 \alpha^i_1 \alpha^{i+1}_1 - { \alpha^\prime }^i_2 \alpha^i_2 \alpha^{i+1}_2 
\right)
\right]
 t_1\: t_2 \: t_3 \: t_4
\end{equation*}
with the following $t_i$ matrices
\begin{eqnarray*}
t_1&=&
\left( \begin{array}{ll}
 H^{\frac{1}{2}}& 0 \\ 
0 & H^{-\frac{1}{2}}
 \end{array}\right) 
\otimes
\left( \begin{array}{ll}
 {H^*}^{\frac{1}{2}}& 0 \\ 
0 & {H^*}^{-\frac{1}{2}}
 \end{array}\right), 
\\
t_2&=&
\left( \begin{array}{ll}
 A^{\frac{1}{2}\alpha^i_1}& 0 \\ 
0 & A^{-\frac{1}{2}\alpha^i_1}
 \end{array}\right) 
\otimes
\left( \begin{array}{ll}
 {A^*}^{\frac{1}{2}\alpha^i_2}& 0 \\ 
0 & {A^*}^{-\frac{1}{2}\alpha^i_2}
 \end{array}\right),\\
t_3&=&
\sum_{\beta^\prime=\pm 1}\left( \begin{array}{ll}
 A^{\frac{1}{2}\beta^\prime}& 0 \\ 
0 & A^{-\frac{1}{2}\beta^\prime}
 \end{array}\right) 
\otimes
\left( \begin{array}{ll}
 {A^*}^{\frac{1}{2}\beta^\prime}& 0 \\ 
0 & {A^*}^{-\frac{1}{2}\beta^\prime}
 \end{array}\right),
\\
t_4&=&
\sum_{\beta=\pm 1}\left( \begin{array}{ll}
A^{\frac{1}{2}\beta}& A^{-\frac{1}{2}\beta} \\ 
A^{-\frac{1}{2}\beta} & A^{\frac{1}{2}\beta^\prime}
 \end{array}\right) 
\otimes
\left( \begin{array}{ll}
 {A^*}^{\frac{1}{2}\beta}& {A^*}^{-\frac{1}{2}\beta} \\ 
{A^*}^{-\frac{1}{2}\beta} & {A^*}^{\frac{1}{2}\beta}
 \end{array}\right). 
\end{eqnarray*}

Using the trace identity above to bring together the degrees of freedom via the tensor product of matrices is the main strategy for calculating the Renyi entropy. These calculations look tedious but are greatly simplified using Mathematica. The next step is to bring together the same degrees of freedom in $R_{1,2}R_{2,1}$ using the trace identity which results in $16\times16$ matrices that only depend on $\alpha$ and $\alpha^\prime$ degrees of freedom for both replicas. Notice that so far we have integrated out all $\beta$ and $\beta^\prime$ degrees of freedom (for both replicas).

Now since the $\alpha^\prime$ degrees of freedom do not interact we can do the sum directly. The $\alpha$ degrees of freedom have an Ising type interaction so the transfer matrix method can be applied again promoting functions of $\alpha$ (which appear in $16\times16$ matrices) to $4\times4$ matrices (because there are $2$ replicas). This operation can be done very simply in Mathematica. At the end of the day the Renyi entropy will be given by the trace of the product of $P$ (the number of boundary plaquettes) $64\times64$ transfer matrices. 

If the boundary of the subsystem is largely smooth, i.e. only has a few sharp corners, the different transfer matrices at the corners will give an order-one contribution to the Renyi entropy and the leading order area law term is
\[
 S_2(\rho_A)\approx-P \log \Lambda
\] 
where $\Lambda$ is the largest eigenvalue of the smooth boundary $64\times64$ transfer matrix constructed above. 

An exact calculation of the entanglement entropy is needed for evaluating the topological entropy. Therefore corrections from concave and convex corners must be taken into account by using different transfer matrices at the corners in the boundary. There is a small complication due to the the fact that a convex (concave) corner plaquette has an additional $\alpha$ ($\beta$) spin. Let us review our method: we first consider two matrix elements each given by the trace of a product of $2\times2$ matrices, we then bring together the same degrees of freedom and explicitly sum over $\beta$s to get the  $4\times4$ transfer matrices. Now having two $\beta$s to sum over does not change anything except we have to do a few more summations to get $T_{4\times4}$. With our method of first integrating out the $B$ spins and then the $A$ spins, the additional $\alpha$ spin at a convex corner gives a jump in the indices of the transfer matrices at the end of the day but this can be easily accounted for by inserting a matrix $C_{ij}=1$ between the transfer matrices and calculating the trace of their product as usual.

Let us now write the $2\times2$ matrices for the corners. First consider the transfer matrix $T^{L>}_{2\times 2}$ for entering a convex corner as seen in the top right corner of Fig.~\ref{fig:transfer_all}. Each transfer matrix is written for a $\beta$ bond and the $\alpha$ spins before ($\alpha_{\rm b}$) and after $\alpha_{\rm a}$ that bond.  As mentioned above the discontinuity in the $\alpha$ index ($\alpha_{\rm a}$ for a transfer matrix is not equal to $\alpha_{\rm b}$ for the next transfer matrix) can be taken care of by inserting the matrix $\openone_{16\times 16}\otimes C_{4\times4}$ ($4\times4$ because we have two replicas) at the end of the day between the $64\times 64$ transfer matrices. For exiting a convex corner plaquette, with the notation of the bottom left corner of Fig.~\ref{fig:transfer_all}, we can write a $2\times 2$ transfer matrix $T^{R>}_{2\times 2}$ and similarly a transfer matrix $T^<_{2\times 2}$ for a concave corner (bottom right corner of Fig.~\ref{fig:transfer_all}). In terms of the spin variables shown on the corresponding corner of Fig.~\ref{fig:transfer_all}, these matrices are given by
\begin{widetext}
\begin{eqnarray*}
T^{L>}_{2\times 2}&=&
A^{ \frac{1}{2} {\alpha^\prime}^i }
\exp \left[ i \mu t \:{\alpha^\prime}^i \beta^i \alpha^i_{\rm b} \alpha^{i}_{\rm a} \right]\times
\left( \begin{array}{ll}
 H^{\frac{1}{2}}A^{\frac{1}{2}({\beta^\prime}^i+\alpha^i_{\rm b})}&0  \\ 
 0&H^{-\frac{1}{2}}A^{-\frac{1}{2}({\beta^\prime}^i+\alpha^i_{\rm b})}
\end{array}
\right)
\left(\begin{array}{ll}
 A^{\frac{1}{2}\beta^i}&A^{-\frac{1}{2}\beta^i}  \\ 
 A^{-\frac{1}{2}\beta^i}&A^{\frac{1}{2}\beta^i}
\end{array}
\right)
\left(\begin{array}{ll}
 A^{\frac{1}{2}\alpha^i_{\rm a}}&0  \\ 
 0&A^{-\frac{1}{2}\alpha^i_{\rm a}}
\end{array}
\right) \nonumber\\
T^{R>}_{2\times 2}&=&
A^{ \frac{1}{2} {\alpha^\prime}^i }
\exp \left[ i \mu t\: {\alpha^\prime}^i \beta^i \alpha^i_{\rm b} \alpha^i_{\rm a} \right] 
\times 
\left( \begin{array}{ll}
 H^{\frac{1}{2}}A^{\frac{1}{2}\alpha^i_{\rm b}}&0  \\ 
 0&H^{-\frac{1}{2}}A^{-\frac{1}{2}\alpha^i_{\rm b}}
\end{array}
\right)
\left(\begin{array}{ll}
 A^{\frac{1}{2}\beta^i}&A^{-\frac{1}{2}\beta^i}  \\ 
 A^{-\frac{1}{2}\beta^i}&A^{\frac{1}{2}\beta^i}
\end{array}
\right)\nonumber\\
T^<_{2\times2}&=&
\exp \left[ {i \mu t\: \alpha^i_{\rm b} \alpha^i_{\rm a} \beta^i \beta^{i+1} } \right]
\times 
\left( \begin{array}{ll}
 H^{\frac{1}{2}}A^{\frac{1}{2}(\beta_i+{\beta^\prime}^i+\alpha^i_{\rm b})}&0  \\ 
 0&H^{-\frac{1}{2}}A^{-\frac{1}{2}(\beta_i+{\beta^\prime}^i+\alpha^i_{\rm b})}
\end{array}
\right)
\left(\begin{array}{ll}
 A^{\frac{1}{2}\beta^{i+1}}&A^{-\frac{1}{2}\beta^{i+1}}  \\ 
 A^{\frac{1}{2}\beta^{i+1}}&A^{-\frac{1}{2}\beta^{i+1}}
\end{array}
\right).
\end{eqnarray*}
\end{widetext}

Notice that for a concave corner, the transfer matrix is written between two neighboring boundary plaquettes which in this case do not have a common edge because the plaquette in the middle lies entirely in subsystem $B$ and is therefore not a boundary plaquette. The procedure for going from the initial $2\times 2$ transfer matrix to the final $64\times 64$ is analogous to the smooth boundary case discussed above and a detailed discussion would not be illuminating. We have explicitly verified using the transfer matrix method described in this appendix that the order-one correction to the area law is entirely from the convex and concave corners and the topological entanglement entropy remains identically zero.

%\begin{figure}
% \centering
% \includegraphics[width =3 cm]{concave}
% \caption{The transfer matrix $ T^3_{2\times2}$ at a concave corner and the corresponding
%spin variables.}
% \label{fig:concave_transfer}
%\end{figure}

%\begin{figure}
% \centering
% \includegraphics[width =3 cm]{convex2}
% \caption{The transfer matrix $ T^2_{2\times2}$ for exiting a convex corner and the corresponding
%spin variables.}
% \label{fig:convex_transfer2}
%\end{figure}

\section{Renyi entropy at order $h^4$}
\label{App_C}

We stated in section~\ref{sec:non-integrable} that Eq.~(\ref{eq:second-order}) gives zero topological entanglement at order $h^2$. Consider the partitions of Fig.~\ref{fig:partitions} with the spins inside (but not on the boundary of) the blue regions belonging to subsystem $B$ and all other spins to $A$. For partition I for example, the outer and inner boundaries of the subsystem $B$ are squares of side $\ell_1$ and $\ell_2$ respectively. The whole system is on  an $\ell \times \ell$ torus. In terms of these dimensions, we can explicitly find the quantities $n_{{\rm p}A}$, $n_{{\rm p}B}$, $n_{{\rm d}A}$ and $n_{{\rm d}B}$ for the four different partitions and we observe that the value of each quantity for partitions II and III, which are obviously equal, is the average of those for partitions I and IV. This immediately implies that the topological entanglement entropy identically vanishes at second order in $h$. For completeness we include the results of this simple geometry exercise for partitions I and IV below 
\begin{eqnarray*}
n^{\rm I}_{{\rm p}A} &=& \ell^2-\ell_1^2+\ell_2^2\\
n^{\rm I}_{{\rm p}B} &=& (\ell_1-2)^2-\ell_2^2-4\ell_2\\
n^{\rm I}_{{\rm d}A} &=& 2 (\ell^2 - \ell_1 - \ell_1^2 - \ell_2 + \ell_2^2)\\
n^{\rm I}_{{\rm d}B} &=&  2 \left( 4 - 5 \ell_1 + \ell_1^2 - 5 \ell_2 - \ell_2^2\right)\\
n^{\rm IV}_{{\rm p}A} &=& \ell^2-\ell_1(\ell_1-\ell_2)\\
n^{\rm IV}_{{\rm p}B} &=& (\ell_1-2)(\ell_1-\ell_2-4)\\
n^{\rm IV}_{{\rm d}A} &=&2 \ell^2 - 2 \ell_1^2 + \ell_2  + \ell_1 (2 \ell_2-3)\\
n^{\rm IV}_{{\rm d}B} &=& 24 + 2 \ell_1^2 + 5 \ell_2 - \ell_1 (15 + 2 \ell_2)
\end{eqnarray*}
where we have assumed $\ell_2-\ell_1>4$.
%partition I:
%\begin{eqnarray*}
%n_{{\rm p}A} &=& \ell^2-\ell_1^2+\ell_2^2\\
%n_{{\rm p}B} &=& (\ell_1-2)^2-\ell_2^2-4\ell_2\\
%n_{{\rm d}A} &=& 2 (\ell^2 - \ell_1 - \ell_1^2 - \ell_2 + \ell_2^2)\\
%n_{{\rm d}B} &=&  
% 2 \left( 4 - 5 \ell_1 + \ell_1^2 - 5 \ell_2 - \ell_2^2\right)
%\end{eqnarray*}
%
%partition II:
%\begin{eqnarray*}
%n_{{\rm p}A} &=& \frac{1}{2}  (2 \ell^2 - 2 \ell_1^2 + \ell_1 \ell_2 + \ell_2^2)\\
%n_{{\rm p}B} &=& 2 + \frac{1}{2}(\ell_1-\ell_2-4)(2 \ell_1+\ell_2-2)\\
%n_{{\rm d}A} &=& \frac{1}{2}(4 \ell^2 - 5 \ell_1 - 4 \ell_1^2 - \ell_2 + 2 \ell_1 \ell_2 + 2 \ell_2^2)\\
%n_{{\rm d}B} &=& \frac{1}{2}(32 + 4 \ell_1^2 - 5 \ell_2 - 2 \ell_2^2 - 25 \ell_1 -2 \ell_1 \ell_2 ))
%\end{eqnarray*}
%
%partition III:
%\begin{eqnarray*}
%n_{{\rm p}A} &=& \ell^2-\ell_1(\ell_1-\ell_2)\\
%n_{{\rm p}B} &=& (\ell_1-2)(\ell_1-\ell_2-4)\\
%n_{{\rm d}A} &=&2 \ell^2 - 2 \ell_1^2 + \ell_2  + \ell_1 (2 \ell_2-3)\\
%n_{{\rm d}B} &=& 24 + 2 \ell_1^2 + 5 \ell_2 - \ell_1 (15 + 2 \ell_2)
%\end{eqnarray*}

\begin{figure}
 \centering
 \includegraphics[width =5 cm]{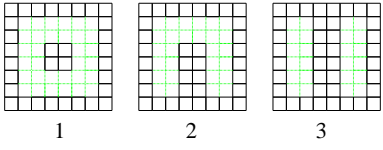}
 \caption{The system for $o(h^4)$ calculation. 
The dashed green bonds belong to subsystem $B$ and the solid black bonds to the subsystem $A$. }
 \label{fig:partition8by8}
\end{figure}

Next, we discuss the fourth order term. The combinatorics get more complicated at higher orders but we can calculate the terms in the expansion by direct enumeration. This gives semi-analytical results with the time-dependence entering analytically through the multiple integrals depending on functions $c$, $p$ and $d$ and the geometry dependence computed for a system of given dimensions by computer enumeration. We have performed these calculation for several systems. Here we present the details for an $8\times 8$ system shown in Fig.~\ref{fig:partition8by8}.

If ${\rm tr}(\rho_A^2)=1+f h^2+g h^4+\cdots$, we have
\[
 S_2=-f h^2-\frac{1}{2}(2g-f^2) h^4+\cdots
\]
Let us assume we have coefficients $f_i$ and $g_i$ for partition $i=1\cdots 3$. We can then write the fourth order term in the topological Renyi entropy as
\begin{eqnarray*}
 S_{\rm topo}{\big |}_{o(h^4)}&=h^4\left[2g_2-g_1-g_3-\frac{1}{2}(2f_2^2-f_1^2-f_3^2)\right]. 
\end{eqnarray*}

To find the fourth order term in $S_{\rm topo}$, in addition to the $f$ coefficients that we found before, we need to calculate the coefficients $g_1$, $g_2$ and $g_3$ for the different partitions. Looking at the structure of the Eq.~(\ref{eq:Renyi-explicit}), we observe that at fourth order we need to insert four $V$s into the four bins $\langle \cdots \rangle\langle \cdots \rangle\langle \cdots \rangle\langle \cdots \rangle$ which give a total of $35$ terms of five different categories. There is one term of type $(1,1,1,1)$ (with one $V$ in each $\langle \cdots \rangle$) and $12$ $(2,1,0,0)$, $6$ $(2,2,0,0)$, $12$  $(3,1,0,0)$ and $4$ $(4,0,0,0)$ terms. First consider the $(1,1,1,1)$ term. We define a matrix ${\cal A}$ such that 
\[
 {\cal A}_{\alpha, \beta}\equiv\langle \alpha\beta|V|0\rangle
\]
and obtain
\[
g_{(1,1,1,1)}={\rm tr}\left[ \left({\cal A} {\cal A}^\dagger \right) ^2\right].
\]
Now since one $V$ term acting on $|0 \rangle$ can only change the state by flipping either one plaquette or one domino, the $|\alpha\beta\rangle$ states with a non-vanishing $ {\cal A}_{\alpha, \beta}$ can be enumerated and the matrix ${\cal A}$ can be constructed. The elements of the matrix ${\cal A}$ contain integrals $(c)$, $(p)$ and $(d)$ depending on which term in $V$ contributes to the matrix element. We then obtain for example the following expression for partition 1 of the $8\times 8$ system
\begin{eqnarray*}
g_{1,(1,1,1,1)}&=&8 \,{\big[}\, 33554432 (c)(c)(c)(c) + 299 (d)(d)(d)(d)\\
&+& 6336 (d)(d)(p)(p)+ 34176 (p)(p)(p)(p)\\
 &+&  196608 (c)(c)(d)(d) + 2359296 (c)(c)(p)(p)\,{\big]}.
\end{eqnarray*}
We can similarly calculate $g_{(1,1,1,1)}$ for other partitions. The next two categories are $(2,1,1,0)$ and $(2,2,0,0)$. The expression for $g_{(2,1,1,0)}$ and $g_{(2,2,0,0)}$ are given below. 
\begin{eqnarray*}
g_{(2,1,1,0)}=2\;\left(\right.
&-&\langle 0 \beta | VV | 0 \rangle 
 \langle 0 | V^\dagger | \alpha \beta \rangle
 \langle \alpha 0 | V | 0 \rangle 
\\
&-&\langle \alpha 0 | VV | 0 \rangle
 \langle 0 | V^\dagger | \alpha \beta \rangle
 \langle 0 \beta | V | 0 \rangle
\\
&+&\left. \langle \alpha \beta | VV | 0 \rangle
 \langle 0 | V^\dagger | 0 \beta \rangle
 \langle 0| V^\dagger | \alpha 0 \rangle+{\rm c.c.}\;\right)
\end{eqnarray*}
and
\begin{eqnarray*}
 g_{(2,2,0,0)}&=&\left(\;
 \langle 0 \beta | VV |0 \rangle
 \langle 0 | V^\dagger V^\dagger |0 \beta \rangle \right.
+\langle \alpha 0| VV |0 \rangle
 \langle 0 | V^\dagger V^\dagger | \alpha 0 \rangle
\\
&+&\left.\langle 0 | VV |0 \rangle
 \langle 0 | VV |0 \rangle + {\rm c.c.}\;\right).   
\end{eqnarray*}

Computing these expressions in terms of the functions $c$, $d$ and $f$ with direct enumeration yields the following expression for the topological Renyi entropy.
\begin{eqnarray*}\label{eq:Renyi-topo}
S_{\rm topo}{\big |}_{o(h^2)}&=&
-128 \:\big{\lbrace} 
 16(d)(p)\left[(p,d)+(d,p)\right]\\
& &+(d)^4-4(d,d)^2
-16\left[(p,d)+(d,p)\right]^2\\
& &-48(p)^4 +256(p)^2(p,p)-320(p,p)^2 
\big{\rbrace} . 
 \end{eqnarray*}

Notice that the last category $g_{(4,0,0,0)}$ is independent of the partitions and does not contribute to the topological entanglement entropy at fourth order in $h$. We can show that the expression in Eq.~(\ref{eq:Renyi-topo}) above identically vanishes using $(p,p)=(p)^2/2$, $(d,d)=(d)^2/2$ and the following integral identity
\[(p,d)+(d,p)=(d)(p).\]

The fourth order calculation also confirms the notion that a growing ring-like (with the expansion order in $h$) region around the system boundary contributes to the entanglement entropy.

\end{document}